\documentclass[11pt,a4paper]{article}
\usepackage[english]{babel}  
\usepackage[latin1]{inputenc} 
\usepackage[colorlinks=true]{hyperref}
\usepackage{latexsym} 
\usepackage{graphicx}  
\usepackage{exscale}
\usepackage{amsmath,amssymb, amsthm}
\usepackage{mathrsfs} 
\usepackage{pdfpages} 

\newcommand{\p}{\phantom}
\newcommand{\f}[2]{\frac{#1}{#2}}
\newcommand{\df}[2]{\frac{\partial #1}{\partial #2}}
\newcommand{\bse}{\begin{subequations}\begin{align}}
\newcommand{\ese}{\end{align}\end{subeqations}}

\newcommand{\noi}{ \noindent}

\title{The regular conducting fluid model for relativistic thermodynamics}
\author{Brandon Carter}
\date{ Contrib. to 5th M. Grossmann meeting, W. Australia, 
August, 1988; reprinted with corrections by Thomas
M\"adler, September 2012. Observatoire  de Paris, 92195 Meudon.
}
\begin{document}

\maketitle
\begin{abstract}
The ``regular'' model presented here can be considered to be the most
natural solution to the problem of constructing the simplest possible
relativistic analogue of the category of classical Fourier--Euler
thermally conducting fluid models as characterised by a pair of
equations of state for just two dependent variables (an equilibrium
density and a conducting scalar). The historically established but
causally unsatisfactory solution to this problem due to Eckart is
shown to be based on an ansatz that is interpretable as postulating a
most unnatural relation between the (particle and entropy) velocities
and their associated momenta, which accounts for the well known bad
behaviour of that model which has recently been  shown to have very
pathological mixed-elliptic-hyperbolic comportments. The newer (and
more elegant) solution of Landau and Lifshitz has a more
mathematically respectable parabolic-hyperbolic comportment, but is
still compatible with a well posed initial value problem only in such
a restricted limit-case such as that of linearised perturbations of
a static background. For mathematically acceptable behaviour under
more general circumstances, and {\it a fortiori} for the physically
motivated requirement of subluminal signal propagation, only strictly
hyperbolic behaviour is acceptable. Attention is drawn here to the
availability of a more modern ``regular'' solution which, unlike those
of Eckart and of Landau and Lifshitz, is fully satisfactory as far as
all these requirements are concerned. This ``regular'' category of
relativistic conducting fluid models arises naturally within a
recently developed variational approach, in which the traditionally
important stress--momentum-energy density tensor is relegated to a
secondary role, while the relevant covariant 4-momentum co-vectors are
instead brought to the fore in such a way as to suggest a simplifying
ansatz that is obviously more natural than those of Eckart and of
Landau and Lifshitz, and automatically takes care of the causality problem.  
\end{abstract}
\section{Introduction: the three current theories}

Until recently, there was no universally accepted solution to the
problem of constructing a physically satisfactory relativistic
analogue of the standard (general purpose) Euler--Fourier theory of a
thermally conducting fluid, even in its simplest version as an eight
component ``two by two'' category of hydrodynamic models in which the
eight independent field components may be considered to be the
space-time coordinate components of a particle number current vector,
$n^\nu$, and an entropy vector, $s^\nu$ say, where a specific model
within the category is fully characterised by just {\it two}
equations of state involving just {\it two} independent variables, the
latter being conveniently considered as an entropy density, $s$ say, and
a conserved particle density, $n$ say, (as defined with respect to a
frame whose appropriate choice will be seen to be crucial for the good
behaviour of the theory) with the dependent variables taken as a
mass-energy density function, $\rho$ say, and a thermal conductivity
scalar $\kappa$ say, as given by a pair of equation of state functions,
$\rho(s, \,n)$ and $\kappa(s,\,n)$, whose form may be considered to be
obtainable completely from (experimental or theoretical) knowledge
only of the thermal equilibrium limit.

The earliest and most widely known proposal leading to a category of
this type was provided by the theory of Eckart \cite{Eckart}, which
has often been accorded \cite{WeinbergBook} the status of ``standard 
textbook solution'' to the problem, despite the fact that it has been 
long recognised \cite{Cattaneo} as being unsatisfactory from the point 
of view of compatibility with relativistic causality. It was long 
thought to share this with its non-relativistic (Euler-Fourier) 
prototype the property of being mixed parabolic-hyperbolic type,
rather than strictly hyperbolic that one would desire not only on 
physical grounds as a prerequisite for relativistic causality, but
also on mathematical grounds as a prerequisite for the existence of 
Cauchy hyper-surfaces admitting a well-posed initial value problem in 
circumstances more general than the strictly static case. However in 
fact, as was already indicated by the work of Glaviano and Raymond 
\cite{Glaviano}, and as has been made particularly clear by the more 
recent stability analysis of Hiscock and Lindblom \cite{Hiscock}, the 
Eckart theory is far worse in so far as it has  the mathematically and 
physically pathological property of being of mixed elliptic-hyperbolic 
type. As such it is incompatible with stable evolution from freely 
specified initial data on a space-like hyper-surface, even in the case 
of small perturbations on a static background, (so that any attempt to 
use it for numerical computations could be expected to lead to disaster). 

Another such ``two by two'' category is provided by the newer theory
of Landau and Lifshitz \cite{LL}, which has long been unduely
neglected, apparently because it was reputed to be ``essentially
equivalent'' (modulo ``unimportant'' correction terms of quadratic or
higher order in deviations from thermal equilibrium)  to a mere
reformulation of the Eckart theory in a new reference system. In fact
however the work of Hiscock and Lindblom has it made clear that the
Landau--Lifshitz theory is in fact a distinct improvement on its
predecessor in so far as it actually does have the marginally hyperbolic
character that was wrongly believed to characterise the Eckart model,
meaning, to be more explicit, that it is of mixed parabolic-hyperbolic
type. It is thus at least mathematically well behaved in the restricted 
context of linearised perturbations on a strictly static background, on 
which it admits the posing of well behaved  initial value problems.

Despite this substantial advantage of mathematical respectability in
this limited sense (in view of which the correction terms by which it
differs from the Eckart model can hardly be dismissed as entirely
unimportant) the Landau--Lifshitz theory still has the seriously
unsatisfactory feature of being incompatible with the strict
hyperbolicity property needed not only for the physical desideratum of
subliminal propagation in accordance with the usual relativistic
causality requirement, but also for the mathematical requirement of
the existence of well behaved Cauchy surfaces (even in the stationary
but non-static case for which parabolic type characteristic elements
orthogonal to the flow would not be integrable). The purport of this
present communication is to convey the message that the
Landau--Lifshitz category should itself now be considered as having
been made obsolete by the more recent  construction \cite{C1} of a
third such ``two by two''  category that is satisfactorily compatible
with physical as well as mathematical causality requirements, and
whose use may therefore safely be recommended for a wide range of
astrophysical applications.  This ``regular'' category arises as the
natural outcome of the a recently developed variational approach
\cite{C2} in which the 4-momentum co-vectors replace the more
traditional stress-momentum-energy tensor as fundamental entities in 
the formulation of the theory. 

Although the time is long overdue for the Eckart theory to be demoted
from standard textbook status that has been perpetuated by Weinberg
\cite{WeinbergBook, WeinbergAPJ}, and others, and to be relegated to
to the rank of historical curiosity, the Landau--Lifshitz theory
remains nevertheless of some genuine mathematical interest as a
distinguished limit case that might even be practically utilisable as
an approximation in very special circumstances. However if one wants a
simple trouble-free model for general purpose use, it is the
``regular'' theory that should be used.

It is to be mentioned, of course, that while it is entirely suitable
of replacing the Eckart model in textbooks for the role of a general
purpose ``off the peg'' conductivity theory, neither the ``regular''
nor any other such simple ``two by two'' category of models involving
only eight independent component variables can compete with the much
more elaborate categories involving fourteen independent variables
that have been  developed by the work of M\"uller \cite{Muller},
Israel \cite{Israel} and Stewart \cite{Stewart}, the extra six
components corresponding to viscous degrees of freedom that are not
taken into account in the simpler models discussed here (and that are
in fact negligible in a wide range of relevant application to problems
such as thermal diffusion in stellar interiors).  While such much more
complicated fourteen components models have the advantage of being
able to fit detailed applications with comparitivity high precision if
enough is known of numerous relevant parameters and equation of state
functions, they have the corresponding disadvantage of being
unnecessary unweildy and uneconomical for many other applications in
which such complicated and detailed information is either undesirable
or unavailable.

\section{The formally identical differential equations of the three
  theories in their standardised from}

Although they were originally set up using different notation schemes
adapted to the different motivational considerations by which the
three (Eckart, Landau--Lifshitz, and ``regular'')  theories referred
to above were first obtained (this is the reason why it took so long
before the difference between the two former was recognised), it is
possible to clarify the relationship between them by converting them
to a ``standardised'' form in which the necessary differential
equations in all the three cases are formally identical, only the
algebraic relations between the quantities involved being different
from one category to another. In all three cases the first of the
primary equations of state functions, $\rho(n,\,s)$ and $\kappa(n,\,s)$,
may be used to derive secondary equation of state functions for
variables $\mu$, $\Theta$, $P$, respectively interpretable  as
chemical potential, temperature and pressure in accordance with
formulae of form that is familiar from the standard thermal
equilibrium theory, namely 
\begin{equation}
\mu=\df{\rho}{n}\;\;,\qquad
\Theta=\df{\rho}{s}\;\;,\qquad 
P=n\mu +s\Theta -\rho\;\;.
\end{equation}
In all three cases it is possible to obtain a basic conductivity
equation in the form that has been made widely familiar by the many
advocates of the Eckart theory, namely
\begin{equation}
\label{eq:eom_Q}
Q^\mu =- \kappa\Big(\gamma^{\mu\nu}\nabla_\nu \Theta +\dot{u}^\mu \Theta\Big)\;\;,
\end{equation}
the remaining differential equations being the entropy creation formula
\begin{equation}
\label{eq:eom_s}
\nabla_\nu s^\nu = \f{Q^\mu Q_\mu}{\kappa \Theta^2}\;\;,
\end{equation}
the usual particle conservation 
\begin{equation}
\label{eq:eom_n}
\nabla_\nu n^\nu = 0\;\;,
\end{equation}
and the stress-momentum-energy (pseudo) conservation law, 
\begin{equation}
\label{eq:eom_T}
\nabla_\nu T^\nu_{\p{\nu}\mu}=0\;\;,
\end{equation}
where the projected space-metric tensor, $\gamma^{\mu\nu}$, and the
acceleration vector, $\dot{u}^\mu$, are defined in terms of the ordinary
(pseudo-Riemanian) space-time metric tensor, $g_{\mu\nu}$, and of a
certain preferred time-like unit vector, $u^\mu$, in the usual manner,
so that one has
\begin{equation}
\gamma^{\mu\nu} = u^\mu u^\nu + g^{\mu\nu}\;\;, \quad
\dot{u}^\mu=u^\nu\nabla_\nu u^\mu\;\;,\quad
u^\nu u_\nu=-1\;\;.
\end{equation}
In all three cases the heat flux vector $Q^\mu$ is defined in terms of
an appropriate heat transport vector $v^\mu$ by an expression of the form
\begin{equation}
Q^\mu = \Theta s v^\mu\;\;, \quad
v^\nu u_\nu = 0\;\;.
\end{equation}
Although it was not done in the original derivation of the two older
theories, it is also possible in all three cases (and greatly helps to
clarify the comparison between them) to express the 
stress-momentum-energy density tensor in the ``canonical'' form
\begin{equation}
T^\nu_{\p{\nu}\mu} = n^\nu \chi_\mu + s^\nu \Theta_\mu + P g^\nu_{\p{\nu}\mu}\;\;,
\end{equation}
in terms of co-vectors, $\chi_\mu$ and $\Theta_\mu$, respectively
interpretable as the ``chemical'' 4-momentum (per conserved
particle) and the ``thermal''  4-momentum (per unit entropy). 

\section{The essential distinct algebraic structural relations or the 
three models}\label{sec:algebraic_relation}

Despite the apparent identity of the foregoing formal equations of
motion of all three theories, they nevertheless differ radically in
their dynamical behaviour, as a result of essential differences
between the remaining purely algebraic structural relations that are
needed to complete the specification of the theories. The foregoing
description can be seen to involve a total of three subsets of each of
eight distinct component variables, so it thus remains a prescription
determining two of these subsets in terms of the third, since
ultimately there should only be eight truly independent variables in
the theory. In the original Eckart theory and in the new ``regular''
theory the most natural choice for the eight independent variables is
the set of components of the two basic current 4-vectors, $n^\mu$ and
$s^\mu$, whereas in the Landau--Lifshitz theory it is a little more
convenient to take the eight independent variables to be the distinct
subset consisting of the two scalars $n$ and $s$, together with the six
components of the unit vector $u^\mu$ and the orthogonal vector
$v^\mu$; the third set of the eight variables that needs to be
determined (and which would themselves be the fundamental ones in a
Clebsch type Legendre transformed reformulation of the perfectly
conducting limit) are the components of the 4-momentum co-vectors,
$\chi_\mu$ and $\Theta_\mu$.

Eight of the sixteen required algebraic component relationships can
conveniently be expressed in a form that (as was the case for the
differential relations of the preceding section) is the same for all
the three cases. To start with the transport velocity vector is always
expressible in form of the (``bulk'' or ``baryonic'' flow) unit vector,
$b^\mu$ say, along the direction of the conserved particle flux,
$n^\nu$, and of the (``caloric'' flow) unit vector, $c^\mu$ say, along
the direction of entropy flux, $s^\mu$, in the form

\begin{equation}
\label{eq:def_v}
v^\mu = \f{b^\mu}{b^\nu u_\nu} - \f{c^\mu}{c^\nu u_\nu}\;\;,
\end{equation} 
where explicitly 
\begin{equation}
b^\mu = \f{n^\nu}{\sqrt{-n^\nu n_\nu}}\;\;, \qquad
c^\mu = \f{s^\nu}{\sqrt{-s^\nu s_\nu}}\;\;.
\end{equation}
It is also possible in all three cases to express the entropy scalar,
$s$, and the effective thermal 4-momentum co-vector, $\Theta_\mu$, in the
form
\begin{equation}
\label{eq:def_s}
s= -s^\nu u_\nu
\end{equation}
and 
\begin{equation}
\label{eq:def_theta_mu}
\Theta_\mu  =\Theta\, u_\mu\;\;.
\end{equation}
To complete the specification of the theories, however, one still
needs eight more algebraic component relations, namely, an analogue of
\eqref{eq:def_v} to fix the (three independent) components of $u^\mu$,
an analogue of \eqref{eq:def_s} to fix the number density scalar, $n$,
and an analogue of \eqref{eq:def_theta_mu} to fix the particle
4-momentum, $\chi_\mu$, and it is at this stage that the distinction
between the theories become apparent. Indeed the historical deviations
of both the Eckart and the Landau--Lifshitz theories were based on
distinct choices of preferred time-like unit vector at the outset. The
Eckart choice, corresponding to the ``conserved particle rest-frame'',
is expressible in the present notation scheme by simply as
\begin{subequations}
\begin{equation}
u^\mu = b^\mu\;\;.
\end{equation}
The Landau--Lifshitz choice, corresponding to the time-like eigenvalue
of the stress-momentum-energy density tensor, is expressible in the
present notation scheme as
\begin{equation}
u^\mu = \f{\mu n^\mu +\Theta s^\mu}{\mu n +\Theta s}\;\;.
\end{equation}
My original deviation of the ``regular'' model used a more covariant
approach, avoiding  excessive reliance on any single ``preferred
rest-frame'' but when this newer theory is translated into the present
notation scheme it turns out that it requires that the preferred unit
vector should correspond to the ``thermal rest-frame'' as determined
by the entropy current, {\it i.e.} one needs to take 
\begin{equation}
\label{eq:def_u_regular}
u^\mu = c^\mu\;\;.
\end{equation}
\end{subequations}

Let us now move on to consider the particle density scalar, $n$. In the
Eckart theory it is defined with respect to the conserved particles'
own rest-frame which in this case corresponds to the preferred
reference system, {\it i.e.} one has 
\begin{subequations}
\begin{equation}
n=-n^\nu b_\nu = -n^\nu u_\nu\;\;.
\end{equation}
In the Landau--Lifshitz theory $n$ is again defined with respect to a
preferred reference system, but it no longer coincides with the
particles' own rest-frame, so one just has
\begin{equation}
n= -n^\nu u_\nu\;\;.
\end{equation} 
On the other hand in the ``regular'' theory it is the particles' own
rest-frame not the preferred reference system that must be chosen,
{\it  i.e. } one needs to take
\begin{equation}
n=-n^\nu b_\nu \;\;.
\end{equation}
\end{subequations}

\begin{figure}[h]
\begin{center}
 \caption{{ \small Illustration of relative position of particle and
 entropy 4-momentum (associated contra-variant vectors marked as
    heavy lines) with respect to the corresponding flow vectors
    (marked as arrows) for 
(a) (aberrant) Eckart model; (b)
    (better) Landau-Lifshitz model; (c) (congenial) regular model}}

\label{fig:momenta}
\includegraphics[scale=0.4]{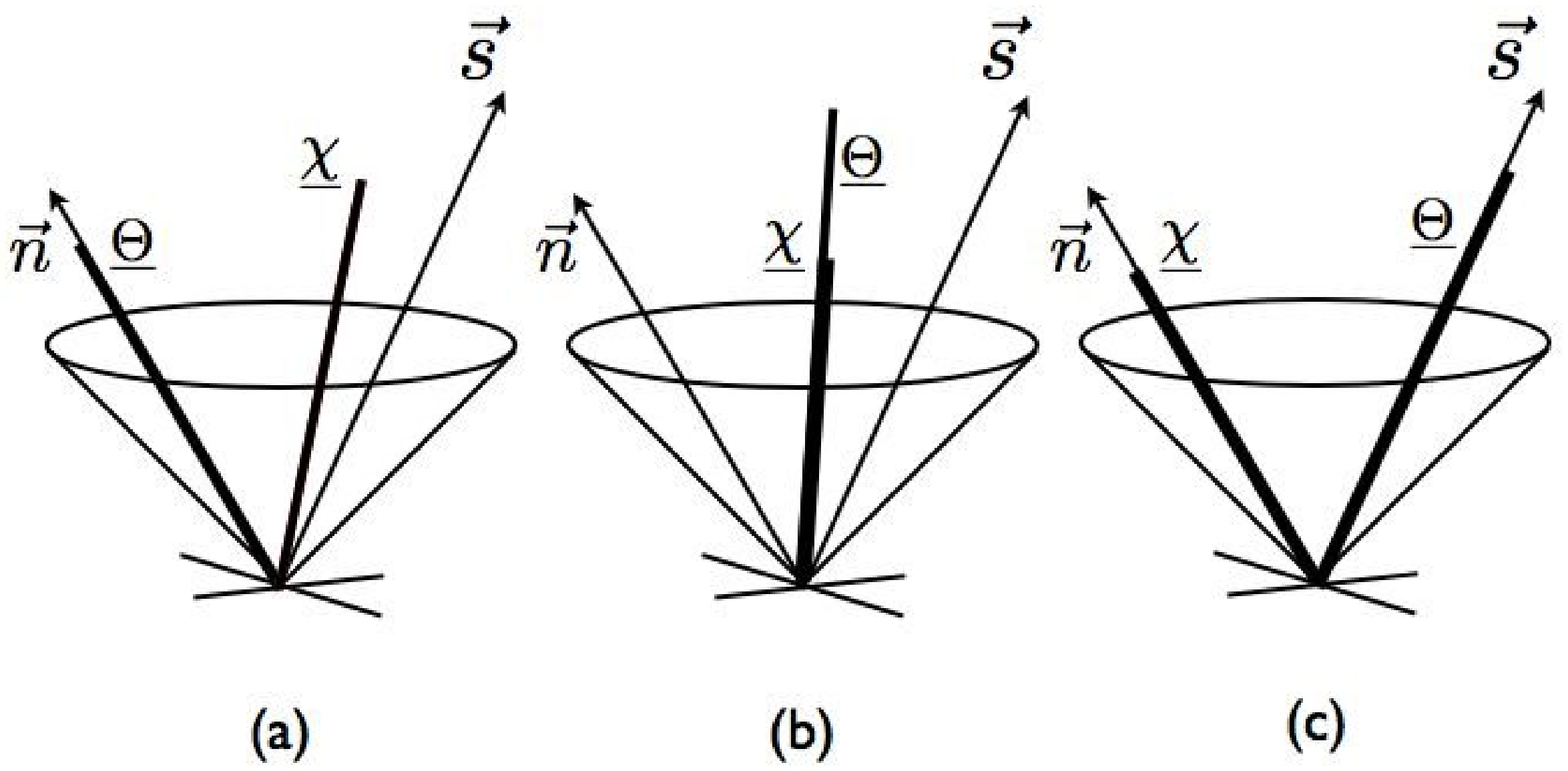}
\end{center}
\end{figure}

Up to the present stage in this presentation there is nothing that
makes it particularly obvious why only the last choice should be fully
satisfactory, why the second is marginally admissible, and why the
first is entirely unacceptable. However the situation becomes much
clearer when we have specified the relation for the particle
4-momentum, $\chi_\mu$ (a concept which  of course was not explicitly
mentioned at all in the original derivations of the two earlier
theories). This final relation, completing the specification of the
model, can at this stage no longer be postulated freely, but is
severely restricted by the requirement that one should avoid an
overdetermination of the system by the equations of motion
\eqref{eq:eom_Q}, \eqref{eq:eom_s}, \eqref{eq:eom_n},
\eqref{eq:eom_T}, which superficially involve nine component equations
of eight unknowns: thus each of the three theories is carefully
contrived so as to ensure that one of these equations --- let's say
the first, {\it i.e.} \eqref{eq:eom_Q}, --- should reduce to an
identity when the others are satisfied. When each theory is fully
specified in accordance with this requirement, the resulting form of
the particle 4-momentum is expressible as follows:

\vspace{1ex}
\noi 
in the Eckart case one has
\begin{subequations}
\begin{equation}
\label{eq:constaint_chi_mu_E}
\chi_\mu = \mu u_\mu +\f{Q_\mu}{n}\;\;;
\end{equation}
in the Landau--Lifshitz case one has simply
\begin{equation}
\label{eq:constaint_chi_mu_LL}
\chi_{\mu}= \mu u_\mu\;\;;
\end{equation}
and finally in the ``regular'' model one has just
\begin{equation}
\label{eq:constaint_chi_mu_reg}
\chi_\mu = \mu b_\mu\;\;.
\end{equation}
\end{subequations}
The basic idea underlying the ``regular'' theory was to apply the most
obviously natural simplifying  ansatz for the 4-momenta, which are
more fundamental than the total stress-momentum-energy density tensor
in the variational approach whose development I have described
elsewhere \cite{C2}: thus the ``regular''  category is distinguished
within a larger category of (in general ``anomalous'') variational
models by the postulate (expressed here by
\eqref{eq:constaint_chi_mu_reg} and \eqref{eq:def_theta_mu} and with
\eqref{eq:def_u_regular} ) that each of the 4-momenta has the same
direction as the (index lowered covariant version of the)
corresponding current.  By contrast the two earlier theories were
based on applying a simplified ansatz (as specified with respect to a
particular preferred reference system) to the total
stress-momentum-energy density tensor:

\noi
in the Eckart case it can be checked that one has 
\begin{subequations}
\begin{equation}
T^\nu_{\p{\nu}\mu} = (\mu n +s\Theta)u^\nu u_\mu +Q^\nu u_\mu + u^\nu
Q_\mu + P g^\nu_{\p{\nu}\mu}\;\;,
\end{equation}
while in the Landau-Lifshitz case one has the even simpler expression
\begin{equation}
\label{ }
T^\nu_{\p{\nu}\mu} = (\mu n +s\Theta)u^\nu u_\mu +P g^\nu_{\p{\nu}\mu}\;\;.
\end{equation}
A comparible version obtained in the ``regular''  case is given by the
expression
\begin{equation}
T^\nu_{\p{\nu}\mu} = \mu n\, b^\nu b_\mu +s\Theta\,c^\nu c_\mu  + P
g^\nu_{\p{\nu}\mu}\;\;.
\end{equation}
\end{subequations}
It is comparatively difficult to analyse and compare the last three
tensorial expressions directly because of the  differing relations
between the reference unit vector $u^\mu$ and the current unit vectors
$b^\mu$ and $c^\mu$. However the simpler co-vectorial expression for
the 4-momenta are much easier to comprehend directly. It is not
necessary to go through the full differential analysis of the
characteristic signal propagation hyper-surface to perceive that there
is something degenerate about the Landau--Lifshitz theory: its
degeneracy is already apparent in the comparison of
\eqref{eq:constaint_chi_mu_LL} with \eqref{eq:def_theta_mu}, whereby
it can be seen that the two momenta are not dynamically independent as
they should be as they are unnaturally constrained to be parallel. The
even more pathological nature of the Eckart theory is also manifest
from \eqref{eq:constaint_chi_mu_E}: quite apparent for its inaesthetic
form, it can be seen to imply that the relative direction of the
momenta in this case is quite unnaturally opposite to the relative
direction of the corresponding currents in this case. It thus
transpires that by unwittingly incorporating such a crazy crossover in
his theory, cf. Figure \ref{fig:momenta}, 
Eckart inadvertently ensured its instability, as an
automatic consequence of a built in negativity of the effective
inertia, which should actually -- for congenial realism -- be positive.

\section{Canonical formulation of regular theory}

In order to convert the ``regular'' theory from the ``standardised''
(Eckart type) formulation we have been using so far back to its
natural ``canonical''  (exterior differential) formulation, in which
it was originally presented \cite{C1, C2} we need to introduce the
entropy transport vector, $\sigma^\mu$, as defined with respect to the
conserved particle (Eckart's preferred) reference frame by

\begin{equation}
\sigma^\mu =s\big(c^\mu + c^\nu b_\nu b^\mu)\;\;,\qquad
c^\nu b_\nu = 0\;\;.
\end{equation}
We also need to introduce the resistivity scalar $Z$ defined by
\begin{equation}
Z= \f{1}{\kappa(b^\mu c_\mu)^2}\;\;.
\end{equation}
In terms of these quantities, the complete set of equations of the
``regular'' model may be written (in a version reminiscent of the four
historic Maxwell equations) as a pair of exterior differential
equations 

\begin{equation}
\label{eq:rot_chi}
2n^\nu \nabla_{[\nu}\chi_{\mu]}+ Z s^\nu \Theta_\nu \sigma_\mu = 0\;\;,
\end{equation}
and 
\begin{equation}
\label{eq:rot_theta}
2s^\nu \Big(\nabla_{[\nu}\Theta_{\mu]}+ Z \sigma_{[\nu} \Theta_{\mu]}\Big) = 0\;\;,
\end{equation}
(where square brackets denote antisymmetrisation of indices), together
with the pair of interior differential equations

\begin{equation}
\label{eq:div}
\nabla_\nu n^\nu=0\;\;, \qquad \nabla_\nu s^\nu = Z \sigma^\nu\sigma_\nu\;\;.
\end{equation}

In this theory it is manifest that (like the Landau-Lifshitz theory,
but unlike the Eckart theory) the ``regular'' theory treats the
conserved particle current and the entropy current on the same footing
in the non-dissipative limit $Z\rightarrow 0$. The form of the
differential equations \eqref{eq:rot_chi}, \eqref{eq:rot_theta},
\eqref{eq:div} holds not only for the ``regular'' category, but but
also more generally for the extended category of (in general
``anomalous'') conducting fluid models of variational type referred to
above \cite{C1, C2}, but in these more general models the algebraic
defining relations will no longer have the simple form for the regular
subcategory in Section \ref{sec:algebraic_relation}.

Editor's note: A review of pertinent progress since this article 
was originally written \cite{C3} has been provided
by Lopez-Monsalo and Andersson \cite{Anders}.

\end{document}